\documentclass{ws-procs975x65}

\begin{document}

\title{Spin Hall Effect in p-type Semiconductors}
\author{SHUICHI MURAKAMI}
\address{Department of Applied Physics, University of Tokyo, 
Tokyo 113-8656, Japan\\
E-mail: murakami@appi.t.u-tokyo.ac.jp}

\maketitle

\abstracts{
The spin Hall effect is a phenomenon of inducing spin current by 
an external electric field. We recently proposed that this
effect can occur in p-type semiconductors without relying upon 
any disorder scattering [S. Murakami {\it et al.}, {\it Science} 
{\bf 301}, 1348 (2003)]. This intrinsic effect is
due to the ``Berry phase in momentum space'', representing topological 
structure of the Bloch band structure.
We explain how the Berry phase brings about the spin Hall effect,
and review several interesting aspects of this effect.}

\section{Introduction}
In the emerging spintronics\cite{prinz1998,wolf2001}
technology, one of the important
issues is to understand the dynamics of spins in metals 
and semiconductors. 
In realizing semiconductor spintronics devices such as Datta-Das spin transistor
\cite{datta1990},
the spin dynamics is vital for effective spin injection/detection,
and spin manipulation.
In this context
one of the recent remarkable findings is the proposals of the intrinsic spin 
Hall effect. In the spin Hall effect, an external electric field 
induces a spin current perpendicular to the field. 
This effect has been considered 
to arise extrinsically, i.e. by impurity scattering.
The scattering becomes spin-dependent in the presence of 
the spin-orbit coupling,
and it results in the spin Hall effect.
In contrast, it has been recently predicted that 
the spin Hall effect can arise 
intrinsically.
It has been proposed for two systems independently, 
one is for n-type semiconductors by the authors
\cite{murakami2003,murakami2003c},
and the other is for two-dimensional p-type semiconductors in 
heterostructure by Sinova {\it et al.}\cite{sinova2003} 
This intrinsic spin Hall effect has theoretically interesting aspects 
similar to the quantum Hall effect,
and it might be useful for effective spin injection 
into semiconductors without using magnetic field.

The spin Hall effect by the electric field along the $l$-axis 
is generally written as 
$j_{i}^{j}=\sigma_{s}\epsilon_{ijl}E_{l}$, where $j_{i}^{j}$ denotes 
the current of the $i$-th component of the spin flowing toward the 
$j$-axis (see Fig.\ \ref{figure1} (a)).
Because both the spin current $j_{i}^{j}$ and the electric field $E_{l}$ 
are even under time-reversal symmetry, the spin Hall conductivity $\sigma_s$
is also even. This represents a reactive (nondissipative) response 
of the system.
This is to be contrasted with usual longitudinal 
charge conductivity $\sigma$; it is odd under time-reversal,
representing a dissipative response.
We also note that because $\sigma_s$ is even under time-reversal,
it can be nonzero even in nonmagnetic materials. 
The criterion for nonzero spin Hall conductivity will be discussed later.

\section{Berry phase in momentum space}
The spin Hall effect can be attributed to ``Berry phase
in momentum space''. Berry phase 
is a geometrical phase associated with an 
adiabatic change of the system\cite{berry1984}. Suppose the Hamiltonian 
depends on some parameter, and the parameter adiabatically changes 
to go back to its initial value at the end. 
The energy levels are assumed to be non-degenerate.
If we assume that the initial state is an eigenstate,
the final state is also the same eigenstate, multiplied by an 
extra phase factor. This phase factor is called the Berry phase.
In many contexts this Berry phase comes into physical phenomena.
For details see Ref.\ \refcite{shaperewilczek-niu}.

In solid state physics, one of the important choices for this parameter
is the wavevector $\mathbf{k}$. This ``Berry phase in momentum space'' 
has been studied in the context of quantum Hall effect and 
also of anomalous Hall effect.
To put it briefly, the intrinsic piece of the Hall conductivity is
determined by a vector field ${\bf B}_{n}({\bf k})$, characterizing
topological structure of the Bloch band. 
This vector field $\mathbf{B}_{n}(\mathbf{k})$ is called a Berry curvature,
and it is defined as
${\bf B}_{ n }({\bf k}) =\nabla_{{\bf k}}\times {\bf A}_{n}({\bf k})$, 
where 
\begin{equation}
A_{n i}({\bf k})=-i\left\langle n{\bf k}\left|\frac{\partial}{\partial
k_{i}}\right|n{\bf k}\right\rangle\equiv -i\int_{{\rm unit\ cell}}
 u_{n{\bf k}}^{\dagger}
\frac{\partial u_{n{\bf k}}}{\partial k_{i}}d^{d}x,
\end{equation}
with $u_{n{\bf k}}({\bf x})$ a periodic part of the 
Bloch wavefunction $\phi_{n{\bf k}}({\bf x})=
e^{i{\bf k}\cdot{\bf x}}u_{n{\bf k}}({\bf x})$.
From the Kubo formula, 
the intrinsic Hall conductivity $\sigma_{xy}$ is expressed 
as 
\begin{equation}
\sigma_{xy} = -\frac{e^{2}}{2\pi h}\sum_{n}\int_{{\rm BZ}}d^{2}k\ n_{F}(
\epsilon_{n}({\bf k})) B_{nz}({\bf k}),
\label{sigmaxy}
\end{equation}
where $n$ is a band index and the integral is over the whole Brillouin zone,
$B_{nz}({\bf k})$ is the $z$-component of the vector
$\mathbf{B}_{n}(\mathbf{k})$, and
$n_{F}(\epsilon_{n}({\bf k}))$ is the Fermi distribution function for 
the $n$-th band.

There is an alternative way to see why this Berry curvature contributes 
to $\sigma_{xy}$.
Sundaram and Niu\cite{sundaram1999}
formulated semiclassical equations of motion including the 
effect of Berry phase. The resulting equations are given by
\begin{eqnarray}
&&\dot{{\bf x}}=\frac{1}{\hbar}
\frac{\partial E_{n}({\bf k})}{\partial {\bf k}}+
{\dot{\bf k}}\times {{\bf B}}_{n}
({\bf k})
\label{dotx}\\
&& \hbar\dot{{\bf k}}=-e({\bf E}+\dot{{\bf x}}\times {\bf B}({\bf x})).
 \label{dotk}
\end{eqnarray}
The second term in Eq.\ (\ref{dotx}) represents an anomalous velocity 
coming from the Berry phase. Without this term, this set of equations
is the conventional one in Boltzmann transport theory.
This anomalous velocity gives rise to the intrinsic 
Hall conductivity as shown below.
First, we note that the current ${\bf J}$ is expressed as
\begin{equation}
{\bf J}=-e\int \frac{d^{d}k}{(2\pi)^{d}}
 \sum_{n}f_{n}({\bf k}){\bf v}_{n}({\bf k}),
\label{J}\end{equation}
where $f_{n}(\mathbf{k})$ is a distribution function.
In the thermal equilibrium ${\bf J}$ vanishes;
a change in $f_{n}({\bf k})$ or ${\bf v}_{n}$ gives rise to nonzero current.
Ohmic current comes from the deviation of $f_{n}({\bf k})$ from 
its equilibrium distribution $n_{F}(\epsilon_{n}({\bf k}))$,
and is carried by electrons near the Fermi surface.
In contrast, the anomalous velocity from the Berry phase 
(the second term in Eq.\ (\ref{dotx})), when plugged into Eq.\ (\ref{J}),
amounts to 
an anomalous current with the following novel properties;
(i) all the occupied states contribute to the current, and 
(ii) the current accompanies no dissipation.
This anomalous velocity is perpendicular to the applied electric field, 
and appear as a Hall effect. 
The resulting formula reproduces 
Eq.\ (\ref{sigmaxy}) obtained by the Kubo formula.

Because of a remarkable similarity between Eqs.\  
(\ref{dotx}), (\ref{dotk}), the Berry curvature ${\bf B}_{n}
({\bf k})=\nabla_{{\bf k}}\times {\bf A}_{n}({\bf k})$ can be regarded as
a ``magnetic field in $\mathbf{k}$-space'', and 
${\bf A}_{n}(\mathbf{k})$ then corresponds to a ``vector potential in 
$\mathbf{k}$-space''. 
We note that this Berry curvature ${\bf B}_{n}({\bf k})$
can have monopoles\cite{fang2003,murakami2003b,berry1984,volovik}.
The monopole density is 
written as
$\nabla_{{\bf k}}\cdot {\bf B}_{n}({\bf k})=2\pi\sum_{l} \nu_{nl}\delta (
{\bf k}-{\bf k}_{nl})$, where $\nu_{nl}$ is an integer.
As ${\bf B}_{n}({\bf k})$ conveys topological properties of the
Bloch wavefunctions, it is natural to ask what corresponds to the 
locations of the monopoles. The answer is the degeneracy points in 
$\mathbf{k}$ space, where 
the $n$-th band touches other bands\cite{berry1984}.
Because the field $\mathbf{B}_{n}(\mathbf{k})$
is radiated from such monopoles, 
the magnitude of $\mathbf{B}_{n}(\mathbf{k})$ 
tends to be larger for $\mathbf{k}$ nearer to 
the monopoles. 
The author recently proposed that 
such monopoles in momentum space affects gap functions in 
magnetic superconductors\cite{murakami2003b}.
First-principle calculations of ${\bf B}_{n}({\bf k})$ 
well explain experimental data of 
anomalous Hall effect\cite{fang2003,yao2003}.

\section{Semiclassical theory of the spin Hall effect in 
p-type semiconductors}
Let us formulate this semiclassical transport theory for 
p-type semiconductors. 
For semiconductors with diamond or zincblende structures, the valence bands
consist of two doubly-degenerate bands, i.e. the light-hole (LH)
and the heavy-hole (HH) bands.
These two bands are described by the following
Luttinger Hamiltonian
\begin{equation}
H=\frac{\hbar^{2}}{2m}\left[\left(\gamma_{1}+\frac{5}{2}\gamma_{2}\right)
k^{2}-2\gamma_{2}({\bf k}\cdot{\bf S})^{2}\right],
\end{equation}
where ${\bf S}=(S_x,\ S_y,\ S_z)$ are spin $3/2$ matrices.
Here we have neglected a tiny contribution coming from 
breaking of inversion symmetry in the zincblende structure.
The helicity $\lambda=\frac{{\bf k}\cdot
{\bf S}}{k}$ is a good quantum number, and it can take 
$\lambda_{H}=\pm\frac{3}{2}$ and $\lambda_{L}=
\pm\frac{1}{2}$ corresponding to the HH and the LH bands, respectively.

\begin{figure}[h]
\begin{center}
\epsfxsize=10cm
\epsfbox{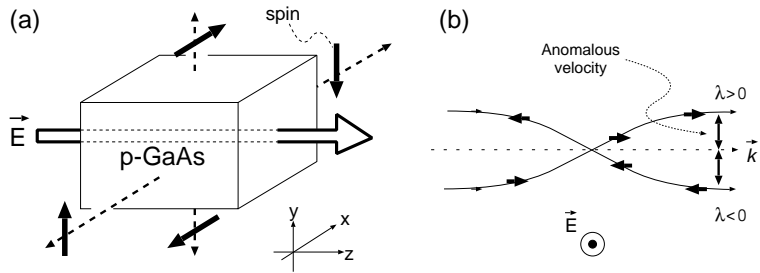}
\caption{(a) Schematic of the spin Hall effect for p-type semiconductors.
(b) Semiclassical trajectories for the wavepackets with 
helicity $\lambda$, projected onto the plane perpendicular to the 
electric field ${\bf E}$. The thick arrows represent the spins.}
\label{figure1}\end{center}
\end{figure}

The semiclassical equations of motion for holes under an electric field 
$\mathbf{E}$ read as 
\begin{equation}
\dot{{\bf x}}=\frac{1}{\hbar}
\frac{\partial E_{\lambda}({\bf k})}{\partial {\bf k}}+
{\dot{\bf k}}\times {{\bf B}}_{\lambda}
({\bf k}),\ \ \ \ \hbar\dot{{\bf k}}=e{\bf E}
\end{equation}
from Eqs.\ (\ref{dotx})(\ref{dotk}). Here $E_{\lambda}(\mathbf{k})=
\frac{\hbar^{2}k^{2}}{2m}(\gamma_{1}+(\frac{5}{2}-2\lambda^{2})\gamma_{2})$
is an eigenenergy for the band with helicity $\lambda$.
The Berry curvature is given by ${\bf B}_{\lambda}
({\bf k})=\lambda(2\lambda^{2}-\frac{7}{2}){\bf k}/k^{3}$,
which depends on the helicity.
This curvature, i.e. the flux density, corresponds to a field
radiated from a monopole at ${\bf k}=0$ with strength
$eg=2\lambda(2\lambda^{2}-\frac{7}{2})$, reflecting that 
the two bands are degenerate at ${\bf k}=0$.
To capture the features of the anomalous velocity from the Berry phase,
we show 
in Fig.\ \ref{figure1} (b) 
the trajectories of holes projected onto a 
plane perpendicular to the electric field\cite{note-relaxation}.
If there is no anomalous velocity, the holes will move
parallel to $\mathbf{k}$. The Berry curvature
gives rise to an anomalous velocity perpendicular to both 
$\mathbf{k}$ and $\mathbf{E}$. This anomalous velocity 
is opposite for opposite helicity, i.e. for opposite spin orientation.
This anomalous velocity from the motion along $\mathbf{k}$, 
when summed over all $\mathbf{k}$ of occupied states, 
amounts to a spin current.
In the zero temperature, by applying an electric field along the 
$l$-direction,
the spin current in which spins along the $i$-axis
will flow along the $j$-direction is calculated as 
\begin{equation}
j_{j}^{i}=\frac{e}{12\pi^{2}}(3k_{F}^{{\rm H}}-k_{F}^{{\rm L}})\epsilon_{ijl}
E_{l}.
\label{spincurrent}
\end{equation}
Here, $k_{F}^{{\rm H}}$ and $k_{F}^{{\rm L}}$ are the Fermi wavenumbers
for the HH and the LH bands, respectively.
A schematic of this effect is shown in Fig.\ \ref{figure1} (a). 

Let us express the above proportionality constant (spin Hall conductivity)
as $\frac{\hbar}{2e}\sigma_{s}$ so that 
$\sigma_{s}$ shares the same unit as the usual charge conductivity 
$\sigma$.
In GaAs at room temperature,
a hole density $n=10^{19}{\rm cm}^{-3}$ yields
$\sigma_{s}=80\Omega^{-1}{\rm cm}^{-1}\cong \sigma $, and
$n=10^{16}{\rm cm}^{-3}$ yields
$\sigma_{s}=7\Omega^{-1}{\rm cm}^{-1}$, and  
$\sigma=0.6\Omega^{-1}{\rm cm}^{-1}$. Thus 
the spin Hall conductivity is as large as or even larger 
than the usual charge conductivity at room temperature.
An energy scale to be compared with the temperature 
is the energy difference between the two valence bands.
If the hole density is as high as $n=10^{19}{\rm cm}^{-3}$,
this energy scale is nominally much larger than room temperature,
and we expect that even at room temperature the spin current 
will remain measurably large.

\section{Physics of the spin Hall effect}
We can discuss a criterion for appearance of the spin Hall effect. 
The spin Hall effect is essentially caused by the spin-orbit coupling.
The spin-orbit coupling brings about a splitting of bands
into multiplets of $\vec{J}=\vec{S}+\vec{L}$.
If all the multiplets split by the spin-orbit coupling 
are filled, the spin Hall effect cancels completely.
Therefore, the criterion for the nonzero spin Hall effect is
a difference of fillings within the same $\vec{J}$-multiplet.
Many kinds of materials, among metals and even band-insulators,
can satisfy this criterion; hence, the spin Hall effect should be quite
a common phenomenon, whereas the size of the effect may vary.

Let us take a few examples to see how this criterion works.
For the valence band $(J=3/2)$ of semiconductors, hole-doping gives rise
to different fillings between the HH and LH bands, and 
the spin Hall effect emerges as the criterion tells us.
On the other hand, the conduction band $(J=1/2)$ 
is doubly degenerate. Therefore, electron-doping does not
give rise to the spin Hall effect. 
In contrast, if we make heterostructures from n-type semiconductors, 
the Rashba term induces the splitting, and results in the spin Hall effect.
Indeed, the spin Hall effect is nonzero as predicted by 
Sinova {\it et al.}\cite{sinova2003}.
Interestingly, the spin Hall effect in this case
takes a universal value $\frac{e}{8\pi}$,
independent of the strength of the Rashba coupling.

So far we assumed an absence of any disorder. 
In reality we cannot escape from disorder and impurities, and 
it is important to clarify the disorder effect. 
There have been several works on this issue, which are mainly on
randomly distributed spinless impurities with a short-ranged potential.
For the p-type semiconductors with spinless impurities, the
self-energy and the vertex correction is calculated within 
the Born and the ladder approximation.
The self-energy turns out to be a constant, and the vertex correction
identically vanishes\cite{murakami2004}. This suggests
that in the clean limit the spin Hall conductivity reproduces the 
intrinsic value calculated earlier\cite{murakami2003}.

In contrast, 
in the 2D n-type semiconductors in heterostructures, it is still
controversial even for spinless impurities. 
Inoue {\it et al.} calculated the self-energy 
and the vertex correction within the Born and ladder approximation 
\cite{inoue2004}.
Remarkably, they showed that with the vertex correction, the spin Hall 
effect becomes zero in the clean limit, as opposed to the 
intrinsic value $\frac{e}{8\pi}$\cite{sinova2003}.
Nevertheless, there are two numerical calculations which contradict
this analytical result. One is based on Kubo formula\cite{nomura2004},
and the other is on the Landauer-B{\"u}ttiker formalism\cite{xiong2004}.
These two calculations suggest that in the clean limit the spin 
Hall conductivity reproduces the intrinsic value $\frac{e}{8\pi}$.
This contradiction remains to be resolved.

We also mention another theoretical issue, concerning a definition of the 
spin current. In the presence of the spin-orbit coupling, as in the 
present problem, the spin is not conserved;
hence, a spin current cannot be uniquely defined. 
In such cases, the spin current is not necessarily defined as 
$\frac{1}{2}\{S_i,\ v_{j}\}$, where $\{\ ,\ \}$ is an anticommutator.
There is no (mathematically) unique definition, 
and it is rather a physics problem as to which definition is
suitable for realistic measurement.
On physical grounds, we have given one reasonable choice for the 
definition of the spin current \cite{murakami2003c}.
For details see Ref.\ \refcite{murakami2003c}.

At present an unambiguous detection of the spin Hall effect has 
still to come. One reason is because detection of spin current is not easy 
in general. A usage of ferromagnetic electrodes for detection would 
bring about a Hall effect from the fringing field, and 
one has to separate it from the spin Hall effect.
On the other hand, the simplest geometry of spin-LED structure does not work
efficiently, because the spins in the induced spin current 
is parallel to the layers\cite{flatte2003}. An ingenious experimental 
setup is 
highly desired.

\section{Conclusion}
In this paper, we briefly review the theoretical predictions 
of the intrinsic spin Hall effect, as a manifestation of the
Berry phase in momentum space. Though there remain some theoretical and 
experimental issues to be clarified, the spin Hall effect 
should consititute an important part in understanding 
the spin dynamics in semiconductors and in implementing
spintronics devices.

\section*{Acknowledgments}
This work is in collaboration with N.\ Nagaosa and S.\ C.\ Zhang, 
and the author would like to thank them for 
fruitful discussions and helpful comments.
This work is supported by 
Grant-in-Aids from the Ministry of 
Education, Culture, Sports, 
Science
and Technology of Japan.


\begin{thebibliography}{99}
\bibitem{note-relaxation}
In reality, the holes will be scattered by disorder before they
run a long way along this trajectory.
\end{thebibliography}
\end{document}